\documentclass[12pt]{article}
\usepackage[dvips]{graphicx}

\begin{document}
\title{Low Momentum Classical Mechanics with Effective Quantum Potentials}
\author{
F.~Haas\footnote{ferhaas@exatas.unisinos.br}\\
Universidade do Vale do Rio dos Sinos - UNISINOS \\
Unidade de Exatas e Tecnol\'ogicas \\
Av. Unisinos, 950\\
93022--000 S\~ao Leopoldo, RS, Brazil}
\maketitle
\begin{abstract}
\noindent A recently introduced effective quantum potential theory is studied in a low momentum region of phase space. This low momentum approximation is used to show that the new effective quantum potential induces a space-dependent mass and a smoothed potential both of them constructed from the classical potential. The exact solution of the approximated theory in one spatial dimension is found. The concept of effective transmission and reflection coefficients for effective quantum potentials is proposed and discussed in comparison with an analogous quantum statistical mixture problem. The results are applied to the case of a square barrier. 
\end{abstract}

\hspace{.25cm} {{\it PACS numbers: 05.60.Gg, 03.65.Nk, 03.65.Xp}}

\section{Introduction}

In nano-MOSFETs, resonant tunneling diodes and other micro-electronic devices, quantum effects are known to play a significant role \cite{Markowich}. Recently, there has been some efforts to create models incorporating quantum effects for micro-electronic devices in a less expensive way than, for instance, the Wigner-Poisson model \cite{Haas}. One of these simplified approaches is based on effective quantum potentials \cite{Feynman}. The aim of effective quantum potential methods is to replace the quantum transport equations by a classical transport theory with a modified potential, reproducing to some extent the essential quantum effects of the problem. To be faithful, an effective quantum potential theory must have similar or lower computational complexity than the quantum transport equations. In addition, it must reduce to classical transport theory for vanishing Planck's constant. Apart from more pragmatic purposes, effective quantum potentials have also attracted attention as an attempt to return to phase space and classical statistical mechanics. For a general introduction on the subject of effective potentials, the reader can see the works \cite{Ferry1, Ringhofer} as well as the references therein. 

Recently \cite{Ringhofer, Ahmed1}, a new effective quantum potential has been proposed and applied \cite{Ahmed2} to model nano-MOSFETs of $25 nm$. The advantage of the new method over the previous ones like  is that it does not require the introduction of a fitting parameter to be adjusted at the beginning of the simulations. Indeed, for effective quantum potentials constructed from Gaussian smoothing of the classical potential, there has been some controversy about the correct value of the smoothing parameter. Specifically \cite{Ferry1}, \cite{Shifren, Ferry2} effective quantum potentials $V_{Gauss}(q)$ obtained from Gaussian smoothing of the classical potential $V^{C}(q)$ according to 
$$
\label{e0}
V_{Gauss}(q) = \frac{1}{\sqrt{2\pi}\,\sigma}\,\int\,V^{C}(q+q')\exp\left(- \frac{{q'}^2}{2\sigma^{\,2}}\right)\,dq' \,, \nonumber
$$
the smoothing parameter being $\sigma$. The  smoothing parameter is related to the nonzero size of the electron wave packet and its value changes in a decisive way the position and the value of the peak of charge density in simulations \cite{Ferry1}. For some candidates for $\sigma$, see reference \cite{Ferry1}. In contrast, in the new effective quantum potential method the size of the wave packet is determined by the particle's energy, with no need of adjustable parameters. The derivation of the effective potential comes from the fact that the Wigner-Bloch and quantum Boltzmann equations should have identical stationary solutions. The details on the derivation can be found in \cite{Ringhofer, Ahmed1}. 

The effective quantum potential introduced in \cite{Ringhofer, Ahmed1} depends in a nontrivial way on momentum. This  obscures its interpretation and the construction of analytical results. From the conceptual viewpoint, this is a disadvantage over methods based on Gaussian smoothing only, where the effective quantum potential is not momentum dependent. To circumvent these difficulties and to obtain a more profound understanding of the new effective quantum potential, the present work proposes a low momentum approximation, res\-tric\-ting to a low momentum region of phase space. As will be seen, for these low energy cases the new effective potential amounts to the introduction of a space-dependent mass and of a smeared potential. For simplicity, we restrict to one-dimensional problems. In this case, the low momentum approximation allows for the complete integrability of the equations of motion. From the exact solution of the dynamics we can get detailed information about the structure of the effective quantum potential theory. 

One of the main objectives of effective quantum potentials is the description of tunneling over barriers using a classical language. Nevertheless, tunneling is a nonlocal phenomena which probably can not be exactly matched by local effective potential theories \cite{Ferry1}. In addition, of course it is somewhat misleading to talk about transmission coefficients and so on when using effective quantum potentials. Indeed, in this case we would have simply ``tunneling" as a result of particles undergoing classical transport over the barrier. Nevertheless, it is interesting to consider in detail the transport of particles in our effective quantum model. For this purpose we consider here the specific case of a square barrier, introducing an effective  transmission  coefficient associated to the effective quantum potential. This effective transmission coefficient is analog to the traditional transmission coefficient of quantum mechanics, but has a different nature. We define it as the fraction of particles surpassing the barrier of the effective quantum potential, for a beam of particles going from the left and with equally spaced energies ranging from zero up to the height of the classical barrier. A similar definition is introduced for the effective reflection coefficient. Even if these effective coefficients are not strictly analog to the quantum ones, they are useful for a quantitative analysis of tunneling in effective quantum potential theories.  

The work is organized as follows. In Section 2, the explicit form of the effective quantum potential is introduced. The effective quantum potential is expanded considering small energies, leaving us with an approximated form characterized by a space-dependent mass and a smeared potential. In Section 3, the exact solution for the classical mechanics associated to the approximated effective quantum potential is constructed. With the aid of the analytical results of Sections 2 and 3, in Section 4 the case of a square barrier is analyzed in detail. In particular, we explicitly obtain the effective quantum potential and the effective transmission and reflection coefficients. We compare these coefficients with the transmission and reflection coefficients found in the strictly quantum case. Section 5 is reserved to the conclusions.

\section{Classical Me\-cha\-nics with Ef\-fec\-ti\-ve Quan\-tum Potentials}

Our effective quantum potential theory is described by 
\begin{equation}
\label{e1}
\dot{q} = \frac{\partial\,\varepsilon^{eff}}{\partial p} \,, \quad \dot{p} = -  \frac{\partial\,\varepsilon^{eff}}{\partial q} \,,
\end{equation}
where the effective Hamiltonian is 
\begin{equation}
\label{e2}
\varepsilon^{eff}(q,p) = \frac{p^2}{2m} + V^{eff}(q,p) \,.
\end{equation}
In equation (\ref{e2}), the effective quantum potential is not simply a Gaussian smoothing, being defined according to the recent works \cite{Ringhofer, Ahmed1},  
\begin{equation}
\label{e3}
V^{eff}(q,p) = \int\,\Gamma(q-q',p)\,V^{C}(q')\,dq' \,,
\end{equation}
where $V^{C}(q)$ is the classical potential and 
\begin{equation}
\label{e4}
\Gamma(q-q',p) = \frac{1}{2\pi}\int\frac{2m}{\beta\hbar\,p\,k}\,\sinh(\frac{\beta\hbar\,p\,k}{2m})\,\exp[-\frac{\beta\hbar^{2}k^2}{8m} + ik(q-q')]\,dk \,.
\end{equation}
In the last equation, $\beta = (\kappa_{B}T)^{-1}$ as usual. Observe that the integral smoothing (\ref{e3}) reduce fluctuations, a welcomed result especially in the neighborhood of hetero-junctions. 

From (\ref{e3}) it follows that the effective quantum potential is momentum dependent. This fact renders the equations of motion difficult to solve in general. However, the form of the integral (\ref{e4}) suggests a useful approximation. Indeed, let us restrict to small momenta, 
\begin{equation}
\label{e5}
\beta\hbar\,pk/m \ll 1 \,,
\end{equation}
to expand the hyperbolic sine at (\ref{e4}). Of course, the approximation (\ref{e5}) has to be carefully justified since $k$ is not fixed in the integral, so that, even for small $p$, the condition (\ref{e5}) may be not satisfied. However, the exponential in (\ref{e4}) is negligible for large $k$, provided
\begin{equation}
\label{e6}
\beta\hbar^{2}k^{2}/m \gg 1 \,.
\end{equation}
When the argument of the hyperbolic sine at (\ref{e4}) is not negligible, so that $\beta\hbar\,pk/m \approx 1$, equation (\ref{e6}) can be used to disregard higher order terms provided
\begin{equation}
\label{e7}
\beta\,p^{2}/m \ll 1 \,.
\end{equation}
Indeed, substituting $\hbar\,k \approx m/(\beta\,p)$ into (\ref{e6}), we get (\ref{e7}). 
For even large wave numbers the approximation becomes still more accurate. In view of the above arguments, for not too large momenta it is justifiable to expand the hyperbolic sine at (\ref{e4}) up to $O(\beta\,p^{2}/m)$. This gives 
\begin{equation}
\label{e8}
V^{eff}(q,p) = V(q) + \frac{1}{2}\left(\frac{1}{M(q)} - \frac{1}{m}\right)\,p^2 + O\left(\left(\frac{\beta\,p^{2}}{m}\right)^{2}\right) \,,
\end{equation}
where
\begin{eqnarray}
\label{e9}
V(q) &=& \frac{1}{2\pi}\int\,dk\,dq'\,\exp[-\frac{\beta\hbar^{2}k^2}{8m} + ik(q-q')]\,V^{C}(q') \,,\\
\label{e10}
M^{-1}(q)\!\!\!\!&=&\!\!\!\!m^{-1}+\frac{\beta^{2}\hbar^2}{24\pi m^2}\!\int\!\!dk\,dq'k^{2}\exp[-\frac{\beta\hbar^{2}k^2}{8m}\!+\!ik(q\!-\!q')]V^{C}(q') \,. \end{eqnarray}

Notice that the expansion is to all orders of $\hbar$, the only restriction being to consider small momenta. Taken into account (\ref{e8}), we obtain the approximated effective quantum Hamiltonian 
\begin{equation}
\label{e11}
\varepsilon^{eff}_{approx} = \frac{p^2}{2M(q)} + V(q) \,,
\end{equation}
the Hamiltonian for a particle of position-dependent mass under a time-independent potential. Quantum corrections are present both in $V(q)$ and $M(q)$. 

\section{Exact Solution}

Hamilton equations for $\varepsilon^{eff}_{approx}(q,p)$ read
\begin{equation}
\label{e12}
\dot{q} = \frac{p}{M} \,, \quad \dot{p} = - \frac{dV}{dq} + \frac{1}{2M^2}\,\frac{dM}{dq}\,p^2 \,.
\end{equation}

Eliminating $p$ and restricting to level sets of constant energy $\varepsilon_{approx}^{eff} = \varepsilon$, the result is 
\begin{equation}
\label{e13}
\ddot{q} = - \frac{1}{M}\,\frac{dV}{dq} + \frac{(V-\varepsilon)}{M^2}\,\frac{dM}{dq} \,,
\end{equation}
or
\begin{equation}
\label{e14}
m\ddot{q} = - \frac{dV^{Q}(q)}{dq} \,,
\end{equation}
with the potential
\begin{equation}
\label{e15}
V^{Q}(q) = \frac{m}{M(q)}\,(V(q) - \varepsilon) + \varepsilon \,.
\end{equation}
The interpretation of (\ref{e14}) is as follows. At least for low momenta, the effective quantum potential induces modifications of the original classical potential, so that the final result is classical motion under the potential (\ref{e15}). The exact solution for (\ref{e14}) then follows from elementary methods. For $\hbar \rightarrow 0$, we have $M \rightarrow m$ and $V \rightarrow V^{C}$, so that $V^{Q} \rightarrow V^{C}$ and the classical Newtonian equation is recovered. The addition of the irrelevant numerical constant $\varepsilon$ at the end of (\ref{e15}) was just a matter of convenience to obtain the correct classical limit when $\hbar \rightarrow 0$.

The usefulness of the form (\ref{e15}) together with (\ref{e9}-\ref{e10}) is that it provides a clean way to observe the modifications produced by quantum effects in the effective quantum potential model. For a given classical potential, we have a recipe for constructing the potential $V^Q$ and to study the corresponding completely integrable Newtonian equation. In particular, this strategy can be used to obtain conclusions about tunneling rates related to the smearing of the classical potential when replaced by the  potential $V^{Q}$. This possibility will be explored in the next Section.

\section{Tunneling}
Consider a square potential barrier centered at $q = 0$, height $V_0$ and width $2L$, given by 
\begin{equation}
\label{e17} 
V^C = V_{0}\,[\theta(L+q) + \theta(L-q) - 1] \,,
\end{equation}
where $\theta$ is the unit step function. The potential (\ref{e17}) was chosen for its simplicity only; the specific choice of the form of the potential barrier does not affect in a decisive way the results. Using (\ref{e9}-\ref{e10}), we obtain 
\begin{eqnarray}
\label{e18}
V &=& \frac{V_0}{2}\left(erf\left(\frac{\sqrt{2m}(L+q)}{\sqrt{\beta\hbar^{2}}}\right) + erf\left(\frac{\sqrt{2m}(L-q)}{\sqrt{\beta\hbar^{2}}}\right)\right) \,,\\
M^{-1} &=& m^{-1} + \frac{2V_0}{3\hbar}\left(\frac{2\beta}{\pi\,m}\right)^{1/2}
\exp\left(-\frac{2m}{\beta\hbar^{2}}(L^2 + q^2)\right) \times \nonumber \\
\label{e19}
&\times& \,\left(L\cosh\left(\frac{4mLq}{\beta\hbar^2}\right)  - q\sinh\left(\frac{4mLq}{\beta\hbar^2}\right)\right) \,,
\end{eqnarray}
where $erf$ is the error function. The quantum potential $V^Q$ then follows from (\ref{e15}). 

Graphics of the space-dependent mass are shown in figure 1, with $m = 1$, $L = 1/2$, $V_0 = 1$, $\beta = 1/8$ and two different values of $\hbar$, namely $\hbar = 10$ (full line) and $\hbar = 30$ (dotted line), using arbitrary units.  As expected, $M \rightarrow m$ for $|q| \gg L$. However, in the neighborhood of the barrier there is a small increase of the mass. In contrast, inside the barrier the mass is lowered. For larger quantum effects, we observe a displacement of the location of the mass peak, as well as an increase of the mass peak. 

In figure 2, we consider the potential $V^Q$ for $m = 1$, $L = 1/2$, $V_0 = 1$, $\beta = 1/8$, $\varepsilon = .25$ and for varying $\hbar$. For $\hbar = 0$, the classical square barrier is recovered. For larger quantum effects ($\hbar = 3$ for the dashed line and $\hbar = 6$ for the dotted line), we see an increasing smoothing of the potential $V^Q$. The height of $V^Q$ becomes smaller for larger quantum effects, increasing tunneling. The observed smoothing is somewhat reminiscent of the smearing of the self-consistent hole potential for quantum plasmas described by the Wigner-Poisson system \cite{Luque}. In spite of the small increase of the mass $M(q)$ near the barrier, the net result of the effective quantum potential is clearly the lowering of the barrier. 

To obtain a condition for tunneling, consider that, for not too big quantum effects, the maximum of $V^Q$ is obtained for $q = 0$. For very large quantum effects, we can show graphically that the maximum of $V^Q$ is not at $q = 0$ anymore, being displaced to symmetric positions around the origin. However, in these cases the smoothing is very strong. Disregarding this possibility, we obtain the analytical expression
\begin{eqnarray}
max(V^{Q}) = V^{Q}(0) &=& \left(1 + \frac{2V_{0}L}{3\hbar}\left(\frac{2\beta\,m}{\pi}\right)^{1/2}
\exp\left(-\frac{2mL^2}{\beta\hbar^{2}}\right)\right) \times \nonumber \\ \label{e20} &\times& \left(V_{0}\,erf\left(\frac{\sqrt{2m}L}{\sqrt{\beta\hbar^{2}}}\right) - \varepsilon\right) + \varepsilon \, 
\end{eqnarray}
For a classical particle of energy $\varepsilon$ to surpass the barrier $V^{Q}$, it is necessary that $\varepsilon > max(V^{Q})$. With (\ref{e20}), this gives 
\begin{equation}
\label{e21}
\varepsilon > V_{0}\,erf\left(\frac{\sqrt{2m}L}{\sqrt{\beta\hbar^{2}}}\right) \,,
\end{equation}
the tunneling condition in our classical description. For $\hbar \rightarrow 0$ it reduces simply to $\varepsilon > V_0$ as expected. Also, tunneling is increased for small temperatures and for a small thickness of the barrier. 

As said before, in the present classical model for a quantum process the word ``tunneling" is to be used with caution. However, we can even assign an effective ``transmission coefficient" to be calculated with the smeared out potential $V^Q$. Indeed, let a beam of classical particles going from the left against the barrier $V^C$. Suppose that the beam is constituted by particles with energies ranging from $0$ to $V_0$, with a homogeneous distribution in energy. In other words, the probability to find a particle with energy in a range $\delta\,E$ is simply $\delta\,E/V_0$. Classically, no one of the particles of this hypothetical beam would surpass the barrier. However, if we admit the model of the effective quantum potential (\ref{e3}) and restrict to small energies, so that $\beta\,V_0 \ll 1$, we can apply the approximation of the preceding sections and define an effective transmission coefficient $t$ given by the ratio of the number of particles that surpass the barrier to those that are reflected. In view of (\ref{e21}), this effective transmission coefficient is 
\begin{equation}
\label{e22}
t = \frac{V_0 - V_{0}\,erf\left(\frac{\sqrt{2m}L}{\sqrt{\beta\hbar^{2}}}\right)}{V_0} = 1 - \,erf\left(\frac{\sqrt{2m}L}{\sqrt{\beta\hbar^{2}}}\right) \,.
\end{equation}
Similarly, we obtain an effective reflection coefficient given by 
\begin{equation}
\label{e23}
r = erf\left(\frac{\sqrt{2m}L}{\sqrt{\beta\hbar^{2}}}\right) \,.
\end{equation}
Of course we have $t + r = 1$. Graphics of $t$ and $r$ are shown in figure 3, where once again we see that larger quantum effects increase tunneling. In this particular example, quantum effects are measured by the non dimensional quantity $H = \sqrt{\beta/m}\,\,\hbar/L$. Significant tunneling begins to happen for $H \approx 2$. For $H = 2.97$ we have already $50 \%$ of the particles surpassing the barrier. 

We have defined a transmission coefficient well suited for the classical description we have adopted. It is interesting to compare this with the truly quantum description. The quantum version of the preceding situation would be a statistical mixture of incoming eigen states of well defined energy, $\{\psi(x) = e^{ikx}\}$, where $\psi(x)$ is a particular eigen function on the ensemble. Here, we are considering particles going from the left ($k > 0$) with an energy $E = \hbar^{2}k^{2}/(2m)$. Hence, we have $dE/V_{0} = \hbar^{2}k\,dk/(mV_{0})$, so that the probability of finding a state with wave-number $k$ in the ensemble would be $\rho(k) = \hbar^{2}k_{0}/(mV_{0})$, for $0 \leq k \leq k_0$ and zero otherwise, where $k_0 = \sqrt{2mV_{0}}/\hbar$. We have the proper normalization $\int_{0}^{k_0}\rho(k)\,dk = 1$ as it should be. 

What is the ``transmission coefficient" associated to this statistical mixture? For just one incoming eigen state $e^{ikx}$, any text book on quantum mechanics gives the result 
\begin{equation}
\label{e24}
t_{one}^Q = \frac{k^{2}(k^{2} - k_{0}^2)}{k^4 - k_{0}^{2}k^2 - (k_{0}^{4}/4)\sinh^{2}(2\sqrt{k_{0}^2 - k^2}\,\,L)}
\end{equation}
for this transmission coefficient. For the statistical mixture just described, due to the linearity of the Schr\"odinger equation the transmission coefficient is given by the  average
\begin{equation}
\label{e25}
t^Q = \int_{0}^{\infty}\,t_{one}(k)\,\rho(k)\,dk \,, 
\end{equation}
or, after some easy scalings and rearrangements, 
\begin{equation}
\label{e26}
t^Q = 2 \int_{0}^{1} \frac{x^{3}(x^{2}-1)\,dx}{x^4 - x^2 - (1/4)\sinh^{2}(2\,\sqrt{1-x^2}/Q)} \,.
\end{equation}
In (\ref{e26}) the quantum transmission coefficient is a function of the  parameter $Q = (k_{0}L)^{-1} = \hbar/(\sqrt{mV_{0}}\,\,L)$ only, while the parameter $H = \sqrt{\beta/m}\,\hbar/L$ is present in the effective quantum potential description. In figure 4 we show the quantum transmission coefficient $t^Q$ as well as the quantum reflection coefficient $r^Q = 1 - t^Q$ as functions of $Q$. For $Q = 1.75$ we have already $t^Q = 0.5$. As seen comparing figures 3 and 4, the transmission coefficients for the effective quantum potential and the  quantum statistical mixture problem have similar functional behaviors in spite of the somewhat different analytical expressions (\ref{e22}) and (\ref{e26}) and the different relevant quantum parameters, $H$ and $Q$. In a sense, this supports the viewpoint that $V^Q$ adequately simulates tunneling, even in our low momentum approximation. Other examples (double square barriers, other classical potential forms) can be easily constructed, with results similar to those of this Section. 

\section{Conclusion}

In this work, the effective quantum potential (\ref{e3}) was studied in a low energy region of phase space, characterized by $\beta\,p^{2}/m \ll 1$. In its original form, the effective quantum potential is a complicated function of momentum. In the low momentum approximation, we identified a space-dependent mass and a smoothed potential, both obtainable from the classical potential. For one-dimensional problems, the dynamics of a particle with space-dependent mass under a time-independent potential is completely integrable, a fact that allows for a number of conclusions. In particular, we have found the potential $V^Q$ at equation (\ref{e15}), which, in the low momentum approximation, describes the classical motion under the effective quantum potential. In connection to the problem of a square barrier, we proposed effective transmission and reflection coefficients for effective quantum potentials, which were explicitly found with the aid of the exact solution for the equation of motion. Comparison with true quantum transmission and reflection coefficients is allowed if an appropriated quantum statistical mixture is introduced. In the case of a square barrier, we identified two non dimensional parameters measuring the relevance of quantum effects for tunneling, $H = \sqrt{\beta/m}\,\,\hbar/L$ for the effective quantum potential model and $Q = \hbar/(\sqrt{mV_{0}}\,L)$ for the quantum statistical mixture. 

In conclusion, we have obtained a more detailed physical interpretation of the effective quantum potential (\ref{e3}), in terms of a space-dependent mass and a smoothed potential, in a low momentum approximation. It would be interesting to generalize the results of this work to more dimensions. In particular, it would be interesting to investigate the integrability of potentials $V^{Q}$ obtained from higher dimensional integrable classical potentials.  

\vskip 1cm
\noindent{\bf Acknowledgments}\\
We thanks the Brazilian agency Conselho Nacional de Desenvolvimento Cien\-t\'{\i}\-fi\-co e Tecn\'ologico (CNPq) for financial support.

\newpage
\begin{figure}
\caption{
Space-dependent mass with $m = 1$, $L = 1/2$, $V_0 = 1$, $\beta = 1/8$ and two different values of $\hbar$, namely $\hbar = 10$ (full line) and $\hbar = 30$ (dotted line), using arbitrary units.}
\label{fig1}
\end{figure}
\begin{figure}
\caption{
The potential $V^Q$ for $m = 1$, $L = 1/2$, $V_0 = 1$, $\beta = 1/8$, $\varepsilon = .25$ and $\hbar = 0$ (full line), $\hbar = 3$ (dashed line) and $\hbar = 6$ (dotted line). We consider arbitrary units.}
\label{fig2}
\end{figure}
\begin{figure}
\caption{
Transmission coefficient $t$ (full line) and reflection coefficient $r$ (doted line) as functions of the non dimensional parameter $H = \sqrt{\beta/m}\,\,\hbar/L$.}
\label{fig3}
\end{figure}
\begin{figure}
\caption{
Transmission coefficient $t^Q$ (full line) and reflection coefficient $r^Q$ (doted line) as functions of the non dimensional parameter $Q = \hbar/(\sqrt{mV_{0}}\,L)$.}
\label{fig4}
\end{figure}
\end{document}